\documentclass[preprint,12pt]{revtex4-2}
\usepackage{amsfonts,amssymb,amsmath,epstopdf,epsfig,amsfonts,color,comment}
\usepackage[dvipsnames]{xcolor}
\usepackage{ulem}
\usepackage{comment}
\usepackage{hyperref}
\numberwithin{equation}{section}

\def\dd{\mbox{d}}

\def\e{\epsilon}

\def\D{\Delta}

\begin{document}

\title{On  the Wave Kinetic Equation in the presence of forcing and dissipation}%

\author{D. Maestrini$^1$, D. Noto$^{2,3}$, G. Dematteis$^1$, and M. Onorato$^{1,4}$\\
 ${}^1$ Dipartimento di Fisica, Università di Torino, Via P. Giuria 1, Torino 10125, Italy.\\
 $^2$ Institut Jean Le Rond d’Alembert, Sorbonne Université, Paris, France.\\
 $^3$ CNRS, UMR 9015, LISN, Université Paris-Saclay, Orsay CEDEX 91405, France.\\
 $^4$ INFN, Sezione di Torino, Via P. Giuria 1, Torino 10125, Italy.}
\begin{abstract}
The wave kinetic equation has become an important tool in different fields of physics. In particular, for surface gravity waves, it is the backbone of wave forecasting models. Its derivation is based on the Hamiltonian dynamics of surface gravity waves. Only at the end of the derivation are the non-conservative effects, such as forcing and dissipation, included as additional terms to the collision integral.
In this paper, we present a first attempt to derive the wave kinetic equation when the dissipation/forcing is included in the deterministic dynamics. If, in the dynamical equations, the dissipation/forcing is one order of magnitude smaller than the nonlinear effect, then the classical wave action balance equation is obtained and the kinetic time scale corresponds to the dissipation/forcing time scale.  However, if we assume that the nonlinearity and the dissipation/forcing act on the same dynamical time scale, we find that the dissipation/forcing dominates the dynamics and the resulting collision integral appears in a modified form, at a higher order. 
\end{abstract}

\maketitle
\newpage

\section{Introduction}
Wave turbulence theory (WTT) provides the statistical description of the evolution of the wave action spectral density of a weakly nonlinear interacting system of a large number of waves (\cite{Zakharov2012}, \cite{newell2011wave},\cite{Nazarenko2011}, \cite{galtier2022physics}). The central object of the theory is the wave kinetic equation (WKE), first introduced by \cite{Peierls1929}, which is the analogue for a system of interacting waves of Boltzmann's kinetic equation for particles in a rarefied gas \cite{cercignani1988boltzmann}. For surface gravity waves, the WKE was derived by \cite{hasselmann1962non} and \cite{zakharov1967energy}, and it is currently the building block of wave operational models (\cite{komen1996dynamics}, \cite{cavaleri2007wave}).

Although its rigorous derivation (in a mathematical sense, establishing the convergence  of the asymptotic expansion) has been proved only for the nonlinear Schr\"odinger equation in dimensions larger than one (see \cite{deng2023full,deng2023long}), in the past decades the WKE has been derived (without a rigorous mathematical proof) and widely used in various systems such as internal gravity waves (\cite{olbers1976nonlinear}, \cite{Caillol2000}, \cite{LT2}), capillary waves (\cite{Zakharov1967}), surface waves (\cite{hasselmann1962non,krasitskii1994reduced,Zakharov1999T1234}) plasma waves in magnetohydrodynamics (\cite{Galtier2000}, \cite{Kuznetsov2001}, Bose-Einstein condensation (\cite{Lvov2003}, \cite{Nazarenko2006}), gravitational waves (\cite{Galtier2021}), and vibrating plates (\cite{During2006}, \cite{Cobelli2009}). 
In real systems, dissipation cannot be neglected.
For example, experiments in elastic plates revealed a discrepancy between theoretical predictions on Kolmogorov-Zakharov spectra (\cite{During2006}) and experimental results (\cite{Boudaud2008}, \cite{mordant2010fourier}).
Among all the reasons for which these discrepancies can arise, \cite{Humbert2013} proposed that the origin of the mismatch is due to dissipation, which is inevitably present in all elastic plates.
For surface gravity waves, dissipation due to white-capping or wave breaking also plays an important role in the dynamics, and different models have been developed and phenomenologically included in the energy balance equation (\cite{ardhuin2010semiempirical}, \cite{babanin2011breaking}, \cite{komen1996dynamics}, \cite{liu2019observation}, \cite{cavaleri2007wave}).
In Wave Turbulence Theory, the dissipative effects, as well as the external forcing  are added {\it a posteriori} only after the WKE has been derived for conservative dynamics. However, in principle dissipation and forcing may play a role combined with resonant interactions. Therefore, it is important to build a framework in which dissipation and forcing are taken into account starting from deterministic equations of motion.

In this paper, we consider this possibility and provide a derivation of the WKE, assuming the presence of dissipation/forcing in the deterministic governing equations (in the form of the so-called Zakharov equation).
More specifically, we assume that our system is characterized by two small parameters, $\epsilon$ and $\mu$; the former is related to the strength of the nonlinear interactions and the latter to dissipation/forcing. We emphasize that, although in the literature  $\epsilon$ usually denotes the wave steepness, throughout this manuscript, $\epsilon$ represents the square of the steepness. In this paper, we consider two different scalings between the two parameters: the first is when the dissipation/forcing acts on the kinetic time scale $\mu\sim\epsilon^{2}$, which is the time scale of the four-wave resonances, while the second case of study is when the dissipation/forcing acts on an intermediate time scale $\mu\sim \epsilon$.

\section{The Zakharov equation with dissipation/forcing}\label{sec1}
Throughout the manuscript, we consider a physical domain $[0,L]\times[0,L]$, which implies a two-dimensional infinite discrete Fourier space. The spacing in Fourier space is given by $\Delta k = 2\pi/L$.
We use the following notation
\begin{equation}
\sum_{1234}\equiv\sum_{\mathbf{k}_1,\mathbf{k}_2,\mathbf{k}_3,\mathbf{k}_4},\quad \delta_{12}^{34}\equiv\delta_{\mathbf{k}_1+\mathbf{k}_2,\mathbf{k}_3+\mathbf{k}_4} \enspace \text{ is the Kronecker delta,}\nonumber
    \end{equation}
    \begin{equation}
    X_{i}\equiv X_{\mathbf{k}_i},\qquad \Delta X_{12}^{34}\equiv X_{1}+X_{2}-X_{3}-X_{4}\enspace \text{ for any variable X,}\nonumber
\end{equation}
and the summation goes from $-\infty$ to $+\infty$. It is well known that, under the hypothesis of inviscid and irrotational flow, in the limit of weak nonlinearity, the Euler equations for water waves reduce to the Zakharov equation (\cite{stuhlmeier2024introduction}, \cite{krasitskii1994reduced}).
 In the presence of dissipation, we assume that the Zakharov equation written in terms of the normal variable $a_{\mathbf{k}}$ is corrected by an extra term and takes the form 
\begin{equation}\label{NLSdiss}
i\frac{\dd a_1}{\dd t}=\omega_1a_1+\epsilon\sum_{234} T_{1234}a^*_2a_3a_4\delta_{12}^{34}-i\mu\gamma_1 a_1,
\end{equation}
where $\omega_k=\sqrt{g k \tanh (kh)}$ is the dispersion relation, $k=\vert \mathbf{k} \vert$, and $h$ is the water depth. Although $h$ can be arbitrary, caveats have to be considered in the limit of shallow water in which the system becomes non-dispersive and lacks the natural  randomization of phases, fundamental for the derivation of the WKE. The matrix elements that weight the interactions,  
$T_{1234}$, can be found in \cite{Krasitskii1990} or \cite{Janssen2007}. The term $\gamma_k a_k$ can be interpreted as a forcing or a dissipation: in general, $\gamma_k$ is the sum of two contributions: a positive one that mimics the dissipation and a negative one that plays the role of forcing. The terms $\epsilon$ and $\mu$, both greater than zero, are the  nonlinear and the dissipation/forcing coefficients, 
respectively. To avoid secular growths in the upcoming perturbation theory, we isolate the trivial resonances ${\bf k}_1={\bf k}_2={\bf k}_3={\bf k}_4$; ${\bf k}_1={\bf k}_3$ and ${\bf k}_2={\bf k}_4$; ${\bf k}_1={\bf k}_4$ and ${\bf k}_2={\bf k}_3$ from the nonlinear term in Eq. \eqref{NLSdiss}. We reabsorb them as corrections to the linear oscillation.  The Zakharov equation becomes 
\begin{align}\label{NLSdiss2}
i\frac{\dd a_1}{\dd t}=\Omega_1a_1+\epsilon\sum_{234}' T_{1234}a^*_2a_3a_4\delta_{12}^{34}-i\mu\gamma_1 a_1,
\end{align}
where
\begin{align}\label{ren}
\Omega_1=\omega_1+2\e\sum_{2}T_{1212}|a_2|^2-\e T_{1111}|a_1|^2,
\end{align}
is the renormalized frequency, and the prime in the summation means that only the non trivial interactions are considered. To derive the kinetic equation, we consider the method developed in \cite{Onorato2020};  emphasise that the results that we will obtain do not depend on the method used, and a check has also been carried out using a more standard expansion of the correlators, which confirms our results. 

We introduce two real functions, the action $I_\mathbf{k}$ and the angle $\theta_\mathbf{k}$, such that 
\begin{equation}\label{bk}
a_\mathbf{k}(t)=\sqrt{I_\mathbf{k}(t)}e^{-i\theta_\mathbf{k}(t)}.
\end{equation}
By inserting Eq. \eqref{bk} into Eq. \eqref{NLSdiss2}, separating the imaginary and real parts, the equations for $I_\mathbf{k}$ and $\theta_\mathbf{k}$ read
\begin{align}
\frac{\dd I_1}{\dd t}&=2\epsilon\sum_{234}'T_{1234} \sqrt{I_1I_2I_3I_4}\sin(\Delta \theta_{12}^{34})\delta_{12}^{34}-2\mu \gamma_1 I_1,\label{DIM}\\
\frac{\dd \theta_1}{\dd t}&=\Omega_1+\epsilon\sum_{234}'T_{1234}\sqrt{\frac{I_2I_3I_4}{I_1}}\cos(\Delta \theta_{12}^{34})\delta_{12}^{34},\label{DRE}
\end{align}
with initial conditions $I_1(0)=\bar{I}_1$ and $  \theta_1(0)=\bar{\theta}_1$.

\section{The kinetic equation with dissipation/forcing}
In the Zakharov equation with dissipation/forcing, the dispersion is considered to be a dominant term, whereas nonlinearity and dissipation/forcing are assumed to be small.
In the absence of dissipation/forcing, it is well known that for $\e\ll 1$, the time scale at which the kinetic equation becomes relevant is $t\sim 1/\e^2$. In the Zakharov equation with dissipation/forcing, we have the presence of two small parameters, $\mu$ and $\e$ (which are assumed to be small compared with the dispersive term). Different scalings can be considered: one possibility is that dissipation/forcing acts on the kinetic timescale $\mu\sim\epsilon^2$, and the other case is when dissipation/forcing acts on an intermediate time scale, $\mu\sim\e$. 
In the following, we investigate these two scenarios using the same approach developed in \cite{Onorato2020} with the appropriate modifications due to the presence of dissipation/forcing. (Note that, in Appendix \ref{appE}, we verify our final results using a more conventional approach based on the expansion of the correlators.)

\subsection{The dissipation/forcing acting on the  kinetic time scale \texorpdfstring{ $\mu\sim \epsilon^2$}{epsilon}}
\subsubsection{Small-$\epsilon$ perturbative expansion}
We expand $I_\mathbf{k}$ and $\theta_\mathbf{k}$ in powers of the nonlinear coefficient $\epsilon$ as follows
\begin{align}
I_\mathbf{k}&=I_\mathbf{k}^{(0)}+\epsilon I_\mathbf{k}^{(1)}+\epsilon^2I_\mathbf{k}^{(2)}+\mathcal{O}(\e^3)\label{expI},\\
\theta_\mathbf{k}&=\theta_\mathbf{k}^{(0)}+\epsilon\theta_\mathbf{k}^{(1)}+\epsilon^2\theta_\mathbf{k}^{(2)}+\mathcal{O}(\e^3),\label{expT}
\end{align}
which can be inserted into Eqs. \eqref{DRE} and \eqref{DIM}, and we match powers of $\epsilon$. Because the dissipation/forcing enters at order $\epsilon^2$, the expansion up to order $\epsilon$ remains identical to the one presented in \cite{Onorato2020}. Here, we report the main results: 

\subsubsection*{Order \texorpdfstring{ $\epsilon^0$}{e0}}

At order $\e^0$, we find
\begin{equation}
    \frac{\dd I_1^{(0)}}{\dd t}=0,\qquad \frac{\dd \theta_1^{(0)}}{\dd t}=\Omega_1,\nonumber
\end{equation}
which can be integrated from 0 to $t$, leading to
\begin{equation}\label{sol1}
    I_1^{(0)}(t)=\bar{I}_1,\qquad \theta_1^{(0)}(t)=\bar{\Omega}_1 t+\bar{\theta}_1,
\end{equation}
where the bar denotes the quantity evaluated at time $t=0$. 

\subsubsection*{Order \texorpdfstring{ $\epsilon^1$}{e1}}
At order $\epsilon$, after inserting the results at order $\epsilon^0$, we obtain 
\begin{align}
     \frac{\dd I_1^{(1)}}{\dd t}&=2\sum_{234}' T_{1234}\sqrt{\bar{I}_1\bar{I}_2\bar{I}_3\bar{I}_4}\sin(\Delta \bar{\theta}_{12}^{34}+\Delta\bar{\Omega}_{12}^{34}t)\delta_{12}^{34},\label{dI1}\\
    \frac{\dd \theta_1^{(1)}}{\dd t}&=\sum_{234}'T_{1234}\sqrt{\frac{\bar{I}_2\bar{I}_3\bar{I}_4}{\bar{I}_1}}\cos(\Delta \bar{\theta}_{12}^{34}+\Delta\bar{\Omega}_{12}^{34}t)\delta_{12}^{34}\label{dT1},
\end{align}
which can be integrated from 0 to $t$, leading to
\begin{align}
     I_1^{(1)}(t)&=2\sum_{234}' T_{1234}\sqrt{\bar{I}_1\bar{I}_2\bar{I}_3\bar{I}_4} \frac{\cos(\Delta \bar {\theta}_{12}^{34})-\cos(\Delta \bar {\theta}_{12}^{34}+\Delta \bar {\Omega}_{12}^{34}t )}{\Delta \bar{\Omega}_{12}^{34}}\delta_{12}^{34},\label{I1}\\
    \theta_1^{(1)}(t)&=-\sum_{234}'T_{1234}\sqrt{\frac{\bar{I}_2\bar{I}_3\bar{I}_4}{\bar{I}_1}}\frac{\sin(\Delta \bar {\theta}_{12}^{34})-\sin(\Delta \bar {\theta}_{12}^{34}+\Delta \bar {\Omega}_{12}^{34}t )}{\Delta \bar{\Omega}_{12}^{34}}\delta_{12}^{34}\label{T1}.
\end{align}

\subsubsection*{Order \texorpdfstring{ $\epsilon^2$}{e2}}
After inserting Eqs. \eqref{I1} and \eqref{T1} in the expansions $I_\mathbf{k}(t)=\bar{I}_\mathbf{k}+\epsilon I_\mathbf{k}^{(1)}(t)$ and $\theta_\mathbf{k}(t)=\bar{\theta}_\mathbf{k}+\bar{\Omega}_\mathbf{k}t+\epsilon\theta_\mathbf{k}^{(1)}(t)$, we substitute them into Eq. \eqref{DIM} and  keeping terms at order $\epsilon^2$ gives
\begin{align}
\frac{\dd I_1^{(2)}}{\dd t}=2\sum_{\substack{234\\567}}'\sum_{m=1}^4T_{1234}&T_{m567}\frac{\sqrt{\bar{I}_1\bar{I}_2\bar{I}_3\bar{I}_4\bar{I}_5\bar{I}_6\bar{I}_7}}{\sqrt{\bar{I}_m}\Delta\bar{\Omega}_{m5}^{67}}\delta_{12}^{34}\delta_{m5}^{67}\left\{ \sin(\Delta\bar{\theta}_{12}^{34}+\Delta\bar{\Omega}_{12}^{34}t-\sigma_m \Delta\bar{\theta}_{m5}^{67}) +\right. \nonumber\\
    &\left.\sin\left[\sigma_m(\Delta\bar{\theta}_{m5}^{67}+\Delta\bar{\Omega}_{m5}^{67}t)-\Delta\bar{\theta}_{12}^{34}-\Delta\bar{\Omega}_{12}^{34}t\right]\right\}-2\gamma_1\bar I_1,
\end{align} 
where $\sigma_m=\{+1,+1,-1,-1\}$.
The next step in the derivation of the kinetic equation is to perform averages on the distribution of the initial data.
We assume that phases and actions are independent identically distributed random variables. More specifically, phases will be considered to be uniformly distributed, whereas for actions it is not necessary to specify any distribution (random phase and amplitude (RPA)) assumption, \cite{Nazarenko2011}).
The procedure of averaging over the phases is described in Appendix \eqref{appB}; the first non zero contribution is at second order and is given by 
\begin{align}\label{dI2nonconstant_1}
\frac{\dd \langle I_1\rangle_{\bar{\theta}}}{\dd t}= &4\epsilon^2 \sum_{234}'T_{1234}^2
\bar{I}_1\bar{I}_2\bar{I}_3\bar{I}_4
\sum_{m=1}^4 \frac{\sigma_m}{\bar{I}_m}
\frac{\sin(\Delta\bar{\Omega}_{12}^{34}t)}{\Delta\bar{\Omega}_{12}^{34}}\delta_{12}^{34}- 2\gamma_1\bar I_1.
\end{align}

\subsubsection{The large-box limit}
So far, the calculation has been performed by assuming that in physical space the wave field is periodic in space with period $L$; the derivation of the kinetic equation requires the large box limit, which is achieved by considering the limit $L\to+\infty$, and it corresponds to taking the distance between Fourier modes approaching zero. An additional step in the derivation of the kinetic equation is to evaluate the average over the initial actions which, by hypothesis, are assumed to be statistically independent, i.e. $\langle\bar{I}_1\bar{I}_2\bar{I}_3\bar{I}_4\rangle_{\bar{I}}=\langle\bar{I}_1\rangle_{\bar{I}}\langle\bar{I}_2\rangle_{\bar{I}}\langle\bar{I}_3\rangle_{\bar{I}}\langle\bar{I}_4\rangle_{\bar{I}}$. Therefore, we define the spectral action density 
\begin{align}
   n_\mathbf{k}(t)=n(\mathbf{k},t)=\lim_{\D k\to 0}\frac{ \langle I_k\rangle_{\bar\theta,\bar I}}{(\D k)^2},
\end{align}
and by following the standard rules for correspondence between sums and integrals, Kronecker's delta and the Dirac delta, 
\begin{align}\label{rules}
\sum_{\mathbf{k}}\longrightarrow\int\frac{\dd \mathbf{k}}{(\Delta k)^2},\qquad \delta_{12}^{34}\longrightarrow\delta(\Delta \mathbf{k}_{12}^{34})(\Delta k)^2,
\end{align}
we obtain
\begin{align}\label{dI2nonconstant_3}
\frac{\partial n_1}{\partial \tau_2}=&4\int \dd \mathbf{k}_{234} T_{1234}^2\bar{n}_1\bar{n}_2\bar{n}_3\bar{n}_4\sum_{m=1}^4 \frac{\sigma_m }{\bar{n}_m}\frac{\sin(\Delta\bar{\Omega}_{12}^{34}\tau_2/\epsilon^2)}{\Delta\bar{\Omega}_{12}^{34}}\delta(\Delta \mathbf{k}_{12}^{34})- 2\gamma_1\bar n_1,
\end{align}
where the new time $\tau_2=\epsilon^2t$ has been introduced. 

\subsubsection{The small-\texorpdfstring{$\epsilon$}{e4} limit}
The last step for obtaining the  wave kinetic equation consists of taking the limit for small $\epsilon$.
Recalling that, under the assumption $\tau_2=O(1)$, we have
\begin{align}\label{sincf}
\lim_{\epsilon\rightarrow{0}}
\frac{\sin(\Delta \bar{\Omega}_{12}^{34}\tau_2/{\epsilon^2})}{\Delta \bar{\Omega}_{12}^{34}}=
\pi\delta(\Delta \bar{\Omega}_{12}^{34}),
\end{align}
equation \eqref{dI2nonconstant_3} becomes
\begin{align}\label{dI2nonconstant_4}
\frac{\partial n_1}{\partial \tau_2}=&4\pi\int\dd \mathbf{k}_{234} T_{1234}^2 n_1{n}_2{n}_3{n}_4\left(\frac{1}{n_1}+\frac{1}{n_2}-\frac{1}{n_3}-\frac{1}{n_4}\right)\delta{(\Delta\omega_{12}^{34}})\delta(\Delta \mathbf{k}_{12}^{34})-2\gamma_1 n_1.
\end{align}
Note that we have removed the bar from the right-hand side, assuming that the RPA persists up to the kinetic time.  
This is the kinetic equation with dissipation/forcing that is normally used in wave forecasting models once the functional form of $\gamma_k$ is specified \cite{komen1996dynamics}. A similar equation has been rigorously derived in \cite{Grande2024} for the nonlinear Schr\"odinger equation with an additive stochastic forcing and viscous dissipation.

\subsection{The dissipation/forcing acting on an intermediate time scale \texorpdfstring{$\mu\sim \epsilon$}{e3}}

\subsubsection{Small-$\epsilon$ perturbative expansion}
We now assume that the dissipation takes place on an intermediate time scale, i.e., between the linear and the kinetic one, and we show that a standard perturbation method leads to a secular growth of the solution at order $\epsilon$. In fact, if we start from Eqs. \eqref{DIM} and \eqref{DRE} and we consider the same expansion as in \eqref{expI} and \eqref{expT}, at order $\epsilon^0$ we find again \eqref{sol1}, while at order $\epsilon$ the solution for the action is
\begin{equation}
I_1^{(1)}(t)=2\sum_{234}' T_{1234}\sqrt{\bar{I}_1\bar{I}_2\bar{I}_3\bar{I}_4} \frac{\cos(\Delta \bar {\theta}_{12}^{34})-\cos(\Delta \bar {\theta}_{12}^{34}+\Delta \bar {\Omega}_{12}^{34}t )}{\Delta \bar{\Omega}_{12}^{34}}\delta_{12}^{34}-2 \gamma_1 \bar I_1 t,\label{I1ep}
\end{equation}
which clearly shows a secular growth in time. 
To prevent this behaviour, we start from Eq. \eqref{NLSdiss} and we introduce a new variable 
\begin{align}\label{ItoJ}
J_\mathbf{k}=I_\mathbf{k}e^{2\epsilon\gamma_k t},
\end{align}
so that Eqs. \eqref{DIM} and \eqref{DRE} become
\begin{align}
\frac{\dd J_1}{\dd t}&=2\epsilon\sum_{234}' T_{1234}\sqrt{J_1J_2J_3J_4}\sin(\Delta \theta_{12}^{34})\exp
\left(\e\Delta \gamma_{1}^{234}t\right)\delta_{12}^{34},\label{DIM2}\\
\frac{\dd \theta_1}{\dd t}&=\Omega_1+\epsilon\sum_{234}'T_{1234}\sqrt{\frac{J_2J_3J_4}{J_1}}\cos(\Delta \theta_{12}^{34})\exp
\left(\e\Delta \gamma_{1}^{234}t\right)\delta_{12}^{34}\label{DRE2},
\end{align}
where $\Delta\gamma_{1}^{234}=\gamma_1-\gamma_2-\gamma_3-\gamma_4$ and the initial conditions are $J_1(0)=\bar{J}_1=\bar{I}_1$ and $\theta_1(0)=\bar{\theta}_1$. 
If we expand $J_\mathbf{k}$ and $\theta_\mathbf{k} $ in powers of $\epsilon$, and we match order by order, at order $\e^0$ we obtain:
\begin{equation}\label{sol1b}
    J_1^{(0)}(t)=\bar{J}_1,\qquad \theta_1^{(0)}(t)=\bar{\Omega}_1 t+\bar{\theta}_1.
\end{equation}
\subsubsection*{Order $\e^1$}
Taking Eqs. \eqref{sol1b}, inserting the equations into Eqs. \eqref{DIM2} and \eqref{DRE2}, and keeping terms at order $\epsilon$, leads to  
\begin{align}
\frac{\dd J_1^{(1)}}{\dd t}&=2\sum_{234}' T_{1234}\sqrt{\bar{J}_1\bar{J}_2\bar{J}_3\bar{J}_4}\sin(\Delta \bar{\theta}_{12}^{34}+\Delta\bar{\Omega}_{12}^{34}t)\exp
\left(\e\Delta \gamma_{1}^{234}t\right)\delta_{12}^{34},\label{dI12}\\
\frac{\dd \theta_1^{(1)}}{\dd t}&=\sum_{234}'T_{1234}\sqrt{\frac{\bar{J}_2\bar{J}_3\bar{J}_4}{\bar{J}_1}}\cos(\Delta \bar{\theta}_{12}^{34}+\Delta\bar{\Omega}_{12}^{34}t)\exp
\left(\e\Delta \gamma_{1}^{234}t\right)\delta_{12}^{34}\label{dT12},
\end{align}
whose integrals between $0$ and $t$ are 
\begin{align}
J_1^{(1)}(t)&=2\sum_{234}' T_{1234}\sqrt{\bar{J}_1\bar{J}_2\bar{J}_3\bar{J}_4}\mathcal{P}(\Delta \bar{\theta}_{12}^{34},\Delta\bar{\Omega}_{12}^{34}, \e\Delta\gamma_{1}^{234},t)\delta_{12}^{34},\label{I12}\\
\theta_1^{(1)}(t)&=\sum_{234}'T_{1234}\sqrt{\frac{\bar{J}_2\bar{J}_3\bar{J}_4}{\bar{J}_1}}\mathcal{Q}(\Delta \bar{\theta}_{12}^{34},\Delta\bar{\Omega}_{12}^{34}, \e\Delta\gamma_{1}^{234},t)\delta_{12}^{34}\label{T12},
\end{align}
where we introduced 
\begin{align}
\mathcal{P}(x,y,z,t)&=\frac{e^{z t} [z\sin (x+y t)-y \cos (x+yt)]+y \cos (x)-z \sin(x)}{y^2+z^2},\label{P}\\
\mathcal{Q}(x,y,z,t)&=\frac{e^{z t} [z \cos(x+y t)+y \sin (x+yt)]-y\sin(x)-z\cos(x)}{y^2+z^2}.\label{Q}
\end{align}
Since
\begin{equation}
    \langle\exp(i\Delta \bar{\theta}_{12}^{34})\rangle=\frac{1}{(2\pi)^4}\prod_{m=1}^{4}\int_{0}^{2\pi}\exp(i\sigma_m\bar{\theta}_m)\dd\bar{\theta}_m=0,\nonumber
\end{equation}
the phase averaging of Eq. \eqref{dI12} is zero.

\subsubsection*{Order $\e^2$}
After inserting Eqs. \eqref{I12} and \eqref{T12} into the expansions $J_\mathbf{k}(t)=\bar{J}_\mathbf{k}+\epsilon J_\mathbf{k}^{(1)}(t)$ and $\theta_\mathbf{k}(t)=\bar{\theta}_\mathbf{k}+\bar{\Omega}_\mathbf{k}t+\epsilon\theta_\mathbf{k}^{(1)}(t)$, we substitute them into Eqs. \eqref{DIM2} and  \eqref{DRE2}. Keeping terms at order $\epsilon^2$ leads to
\begin{align}\label{dJdt}
\frac{\dd J_1^{(2)}}{\dd t}&=2\sum_{234}'T_{1234} \sqrt{\bar{J}_1\bar{J}_2\bar{J}_3\bar{J}_4}\left[\frac{1}{2} \left(\frac{J_{1}^{(1)}}{\bar{J}_1}+\frac{J_{2}^{(1)}}{\bar{J}_2}+\frac{J_{3}^{(1)}}{\bar{J}_3}+\frac{J_{4}^{(1)}}{\bar{J}_4}\right)\sin(\Delta \bar{\theta}_{12}^{34}+\Delta\Omega_{12}^{34}t)\right.\nonumber\\
&+\left.\Delta\theta^{(1)34}_{\quad 12}\cos(\Delta\bar{\theta}_{12}^{34}+\Delta\bar{\Omega}_{12}^{34}t)\Biggl]\exp\left(\e\Delta \gamma_{1}^{234}t\right)\right.\delta_{12}^{34},
\end{align} 
where we used $\sin(x+\zeta)\approx \sin(x)+\zeta\cos(x)$ with $x=\Delta\bar{\theta}_{12}^{34}+\Delta\bar{\Omega}_{12}^{34}t$ and $\zeta= \e\Delta\theta^{(1)34}_{\quad 12}$. By using Eqs. \eqref{I12} and \eqref{T12}, after further algebraic manipulations (see Appendix \eqref{appC}), Eq. \eqref{dJdt} becomes
\begin{align}\label{dJ2}
 \frac{\dd J_1^{(2)}}{\dd t}=\sum_{m=1}^4\mathcal{S}_m\frac{R_{1,m}-R_{2,m}}{(\Delta\bar{\Omega}_{m5}^{67})^2+(\e\Delta\gamma_{m}^{567})^2},
\end{align}
where 
\begin{align}
\mathcal{S}_m&=2\sum_{\substack{234\\567}}'T_{1234}T_{m567}\frac{\sqrt{\bar{J}_1\bar{J}_2\bar{J}_3\bar{J}_4\bar{J}_5\bar{J}_6\bar{J}_7}}{\sqrt{\bar{J}_m}}\exp(\e\Delta \gamma_{1}^{234} t)\delta_{12}^{34}\delta_{m5}^{67},\label{Sprime}\\
R_{1,m}&=\sigma_m \e\Delta\gamma_{m}^{567}\mathfrak{Re}\left\{\left(e^{(\e\Delta\gamma_{m}^{567}-i\sigma_m\Delta\bar{\Omega}_{m5}^{67})t}-1\right)e^{i(\Delta\bar{\theta}_{12}^{34}-\sigma_m\Delta\bar{\theta}_{m5}^{67})}e^{i\Delta\bar{\Omega}_{12}^{34}t}\right\},\nonumber\\
R_{2,m}&=\Delta\bar{\Omega}_{m5}^{67}\mathfrak{Im}\left\{\left(e^{(\e\Delta\gamma_{m}^{567}-i\sigma_m\Delta\bar{\Omega}_{m5}^{67})t}-1\right)e^{i(\Delta\bar{\theta}_{12}^{34}-\sigma_m\Delta\bar{\theta}_{m5}^{67})}e^{i\Delta\bar{\Omega}_{12}^{34}t}\right\}.\nonumber
\end{align}
The phase averaging of \eqref{dJ2} (see Appendix \eqref{appC}) is 
\begin{align}\label{dI2nonconstantapp}
\frac{\dd \langle J_1^{(2)}\rangle_{\bar{\theta}}}{\dd t}=&4\sum_{234}'\sum_{m=1}^4T_{1234}^2\bar{J}_1\bar{J}_2\bar{J}_3\bar{J}_4\frac{\sigma_m \Delta\bar{\Omega}_{12}^{34}\sin(\Delta\bar{\Omega}_{12}^{34}t)e^{\e\Delta\gamma_{1}^{234}t}}{\bar{J}_m\left[(\Delta\bar{\Omega}_{12}^{34})^2+(\e p_m)^2\right]}\delta_{12}^{34},
\end{align}
where we defined 
\begin{align}\label{p_m}
p_m=2 \gamma_m-\sum_{j=1}^{4} \gamma_j,
\end{align}
and where we only considered terms proportional to $\e^2$. Note that we have not discarded the $O(\epsilon^2)$ term in the denominator because $\Delta\bar{\Omega}_{12}^{34}$ can be arbitrary small.

\subsubsection{The large-box and small-\texorpdfstring{$\epsilon$}{e5} limits}
As usual, we consider the average over the initial actions, and in the large box limit we define the spectral action density 
\begin{align}
   n_\mathbf{k}^{(J)}(t)=n^{(J)}(\mathbf{k},t)=\lim_{\D k\to 0}\frac{ \langle J_\mathbf{k}\rangle_{\bar\theta,\bar J}}{(\D k)^2}.\nonumber
\end{align}
By following the substitution rules \eqref{rules}, we obtain
\begin{align}\label{dndtau}
\frac{\partial n_1^{(J)}}{\partial \tau_1}=&4\epsilon\int \dd \mathbf{k}_{234}T_{1234}^2\bar{n}_1^{(J)}\bar{n}_2^{(J)}\bar{n}_3^{(J)}\bar{n}_4^{(J)}\sum_{m=1}^4 \frac{\sigma_me^{\Delta\gamma_{1}^{234}\tau_1}\Delta\bar{\Omega}_{12}^{34}\sin(\Delta\bar{\Omega}_{12}^{34}\tau_1/{\epsilon})}{\bar{n}_m^{(J)}\left[(\Delta\bar{\Omega}_{12}^{34})^2+(\epsilon p_m)^2\right]}\delta(\Delta \mathbf{k}_{12}^{34}),
\end{align}
where we have introduced the timescale  $\tau_1=\epsilon t$. 
The last step to obtain the wave kinetic equation consists in taking the limit for small $\epsilon$.
Recalling that 
\begin{equation}
\int_{-\infty}^{+\infty}
\frac{x\sin(x\tau/{\epsilon})}{\left[x^2+(\epsilon p_m)^2\right]}\dd x=\pi e^{-p_m\tau },
\quad \text{and}\quad \lim_{\epsilon\rightarrow{0}}
\frac{\Delta \bar{\Omega}_{12}^{34}\sin(\Delta \bar{\Omega}_{12}^{34}\tau/{\epsilon})}{\left[(\Delta \bar{\Omega}_{12}^{34})^2+(\epsilon p_m)^2\right]}=
\pi e^{-p_m\tau }\delta(\Delta \bar{\Omega}_{12}^{34}),
\end{equation}
Eq. \eqref{dI2nonconstant_3} becomes 
\begin{align}\label{dI2nonconstant_6}
\frac{\partial n_1^{(J)}}{\partial \tau_1}=&4\pi\e\int \dd \mathbf{k}_{234}T_{1234}^2
e^{2\gamma_1 \tau_1}
\bar{n}_1^{(J)}\bar{n}_2^{(J)}\bar{n}_3^{(J)}\bar{n}_4^{(J)}\sum_{j=1}^{4}\frac{e^{-2\gamma_j \tau_1}}{\bar{n}_j^{(J)}}
\delta(\Delta \mathbf{k}_{12}^{34})
\delta(\Delta \bar{\Omega}_{12}^{34}).
\end{align}
Strictly speaking, this equation is valid at initial time. Here, we assume that RPA holds for arbitrary time: thus, we can replace $\bar{n}^{(J)}$ with $n^{(J)}$. Returning to the original variable by using \eqref{ItoJ} 
\begin{equation}
n_\mathbf{k}^{(J)}=n_\mathbf{k}e^{2\gamma_k \tau_1},
\end{equation}
we obtain the final kinetic equation in the dissipative/forced case 
\begin{align}\label{dI2nonconstant_5}
\frac{\partial n_1}{\partial \tau_1}=-2\gamma_1{{n}_1}+&4\pi\e\int\dd \mathbf{k}_{234} T_{1234}^2
n_1n_2n_3n_4 
\sum_{j=1}^{4}\frac{e^{-2p_j \tau_1}}{n_j}\delta(\Delta \mathbf{k}_{12}^{34})
\delta(\Delta{\omega_{12}^{34}}).
\end{align}
Note that the modified collision integral is now a correction to the  dissipative/forced dynamics. 

\begin{figure}
\centering
\includegraphics[scale=0.365]{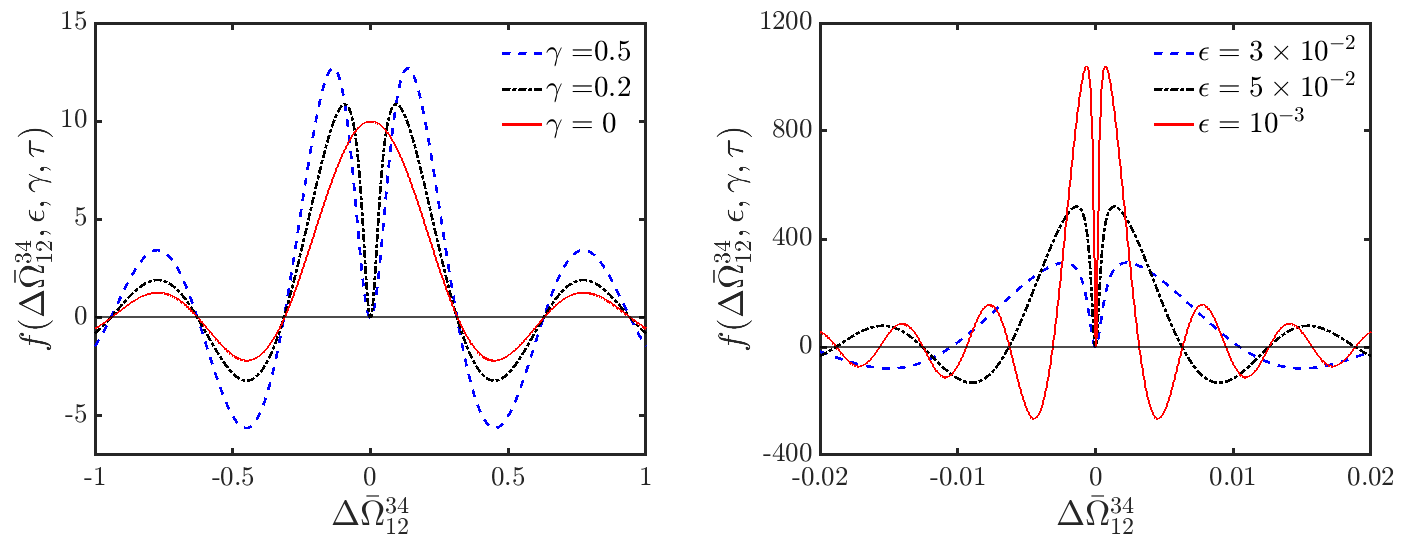}
\caption{On the left, plot of $f(\Delta\bar{\Omega}_{12}^{34},\epsilon,\gamma,\tau)$ for a fixed value of the nonlinearity $\epsilon=0.1$ and for different values of the parameter $\gamma$ as indicated in the legend. The limit $\gamma\to 0$ is the non-dissipative/forced case. On the right, plot of $f(\Delta\bar{\Omega}_{12}^{34},\epsilon,\gamma,\tau)$ as a function of $\Delta\bar{\Omega}_{12}^{34}$ for a fixed value of the dissipation $\gamma=0.1$ and for different values of the parameter $\epsilon$ as indicated in the legend. In both cases, the time is fixed $\tau=1$.}
  \label{fig:fig1}
\end{figure}

\section{Discussion and conclusions}
The energy (or wave action) balance equation is a fundamental tool for operational wave forecasting. Although the nonlinear interactions are obtained directly from the dynamical equations, the forcing and the dissipation source terms are added {\it a posteriori}. The derivation of the wave kinetic equation indicates the time scale at which the collision integral becomes relevant. The time scale associated with the forcing and dissipation in the energy balance equation is more obscure, as those terms are, in general, not derived analytically from first principles. In this paper, we have included a dissipation/forcing term in the dynamical equations and we have attempted a derivation of the wave action balance equation, assuming that the dissipation/forcing is small compared with the dispersion. Two relevant cases are studied: i) the dissipation/forcing in the dynamical equation acts on a longer time scale than the nonlinearity and ii) the dissipation/forcing term acts on the same time scale as the nonlinearity. For the first case, the standard wave action balance equation is derived: the kinetic time scale, $t\sim \epsilon^{-2}$, is the same as the dissipation/forcing time scale. In the second case, the dissipation/forcing dominates the dynamics at a time scale $t\sim \epsilon^{-1}$, and the collision integral appears, as usual, at higher order. However, the collision term is modified by the presence of an exponential term that depends on time. It is interesting to note that if we avoid to take the small $\epsilon$ limit and study the dynamics at fixed $\epsilon$, we are left with a broad function over frequencies that accounts for near resonances, and whose width depends on the dissipation/forcing coefficient. To clarify this issue, if we assume that $\gamma_k=\gamma$, so we can define in \eqref{dndtau} the function
\begin{align}\label{fgamma}
f(\Delta\bar{\Omega}_{12}^{34},\epsilon,\gamma,\tau)=\frac{e^{2\gamma\tau}\Delta\bar{\Omega}_{12}^{34}\sin(\Delta\bar{\Omega}_{12}^{34}\tau/{\epsilon})}{(\Delta\bar{\Omega}_{12}^{34})^2+(2\epsilon \gamma)^2},
\end{align}
which quantifies  near-resonant interactions in the system that occur at finite $\epsilon$. The exactly resonant interactions occurs in the limit as $\epsilon\to 0$ for which $f(\Delta\bar{\Omega}_{12}^{34},\epsilon,\gamma,\tau)$ tends to a function proportional to the Dirac delta. In Fig.\eqref{fig:fig1} on the left we plot the function \eqref{fgamma} by fixing $\e=0.1$, $\tau=1$ for different values of $\gamma$ shown in the legend. 
In the limit as $\gamma\to0$, Eq. \eqref{fgamma} has the same behaviour as the function on the left-hand side of \eqref{sincf} for the non-dissipative/forced case (the difference is a factor $\epsilon$ due to the different time scale). Figure \eqref{fig:fig1} on the right shows the behaviour of \eqref{fgamma} by fixing $\gamma=0.1$, $\tau=1$ for different values of $\epsilon$ reported in the legend. This function has two symmetric peaks about $\bar{\Omega}_{12}^{34}=0$ where its value is zero, and as $\e\to0$, the two peaks
merge at $\bar{\Omega}_{12}^{34}=0$ and the function converges to $\pi e^{2\gamma\tau}\delta(\Delta \bar{\Omega}_{12}^{34})$. From this analysis it is clear that all resonant interactions have zero contribution to the collision integral at short time scale,  as also discussed for conservative systems in \cite{Stiassnie2005}, \cite{janssen03}, \cite{Annenkov2006}. 
Moreover, we note that the envelope of the function in Eq.~\eqref{fgamma} is proportional to the Lorentzian function that has been introduced heuristically in some ``broadened'' versions of the wave kinetic equation (see e.g. \cite{lvov2012resonant}). 

Whether the results presented here will have implications for operational wave forecasting models remains an open question that will be explored in the near future. In the wave forecasting community, it is common to assume that wind input, dissipation due to white-capping, and wave-wave resonant interactions are of the same order of magnitude. We have demonstrated that this description is consistent with a deterministic model, where dissipation and forcing act at a higher order relative to nonlinearity. This assumption allows for the study of these processes in isolation, significantly simplifying the development of wave forecasting systems.
However, it is plausible that, at certain scales, within the deterministic equation of motion, dissipation becomes comparable with the nonlinear interaction term. The transition between these two described regimes is not fully understood, and the derived WKE may provide valuable insights.
As for forcing, there are certainly ocean conditions with strong winds where the forcing term operates on a faster time scale, after which nonlinear interactions could begin transferring energy across the spectrum. Again, our wave kinetic equation could serve as a useful tool for investigating such scenarios.

\textbf{Acknowledgements}: M.O. was funded by Progetti di Ricerca di Interesse Nazionale (PRIN, 2020X4T57A and 2022WKRYNL) and by the Simons Foundation (Award 652354). The authors are grateful to Sergey Nazarenko, Jalal Shatah, and Giorgio Krstulovic for preliminary discussions.


\section{Appendix A}\label{appB}
We define the phase averaging of a function $f(\bar{\theta}_1,\bar{\theta}_2,\dots \bar{\theta}_M)$ over the initial phases $\bar{\theta}_1,\bar{\theta}_2,\dots \bar{\theta}_M$ as
\begin{equation}
\langle f(\bar{\theta}_1,\bar{\theta}_2,\dots \bar{\theta}_M)\rangle_{\bar{\theta}}=\int_{0}^{2\pi}\mathfrak{P}(\bar{\theta}_1,\bar{\theta}_2,\dots \bar{\theta}_M)f(\bar{\theta}_1,\bar{\theta}_2,\dots \bar{\theta}_M)\dd \bar{\theta}_1\dd \bar{\theta}_2\dots\dd \bar{\theta}_M,\nonumber
\end{equation}
where $\mathfrak{P}(\bar{\theta}_1,\bar{\theta}_2,\dots,\bar{\theta}_M)$ is the joint probability density function. If we assume the phases to be statistically independent and uniformly distributed, we have $\mathfrak{P}(\bar{\theta}_1,\bar{\theta}_2,\dots,\bar{\theta}_M)=\mathfrak{P}(\bar{\theta}_1)\mathfrak{P}(\bar{\theta}_2)\dots\mathfrak{P}(\bar{\theta}_M)=\left(2\pi\right)^{-M}$ and hence
\begin{equation}\label{defaverage}
\langle f(\bar{\theta}_1,\bar{\theta}_2,\dots \bar{\theta}_M)\rangle_{\bar{\theta}}=\frac{1}{(2\pi)^M}\int_{0}^{2\pi}f(\bar{\theta}_1,\bar{\theta}_2,\dots \bar{\theta}_M)\dd \bar{\theta}_1\dd \bar{\theta}_2\dots\dd \bar{\theta}_M.
\end{equation}

\section{Appendix B}\label{appC}
We note that   
\begin{align}
\frac{1}{2}\sum_{m=1}^{4}\frac{J_{m}^{(1)}}{\bar{J}_m}&=\sum_{m=1}^{4}\sum_{567}'\sqrt{\frac{\bar{J}_5\bar{J}_6\bar{J}_7}{\bar{J}_m}}\mathcal{P}(\Delta \bar{\theta}_{m5}^{67},\Delta\bar{\Omega}_{m5}^{67},\e\gamma_{m}^{567},t)\delta_{m5}^{67},\label{II1}\\
\Delta\theta^{(1)34}_{\quad 12}=\sum_{m=1}^{4}\sigma_m\theta_m^{(1)}&=\sum_{m=1}^{4}\sum_{567}'\sigma_m\sqrt{\frac{\bar{J}_5\bar{J}_6\bar{J}_7}{\bar{J}_m}}\mathcal{Q}(\Delta \bar{\theta}_{m5}^{67},\Delta\bar{\Omega}_{m5}^{67},\e\gamma_{m}^{567},t)\delta_{m5}^{67},\label{Theta1}
\end{align}
where we used Eqs. \eqref{I1} and \eqref{T1}, and $\sigma_m=\{+1,+1,-1,-1\}$. Plugging Eqs. \eqref{II1} and \eqref{Theta1} into \eqref{dJdt} yields
\begin{align}\label{dI2temp2}
\frac{\dd J_1^{(2)}}{\dd t}&=\sum_{m=1}^{4}\mathcal{S'}_m\left[\mathcal{P}(a_m,b_m,c_m,t)\sin(a+bt)+\sigma_m\mathcal{Q}(a_m,b_m,c_m,t)\cos(a+bt)\right],
\end{align}
where we used Eqs. \eqref{Sprime}, \eqref{P}, and \eqref{Q},  and we introduced 
\begin{align}\label{notation}
a&=\Delta\bar{\theta}_{12}^{34},\qquad b=\Delta\bar{\Omega}_{12}^{34},\qquad 
c=\e\Delta \gamma_{1}^{234},\nonumber\\  a_m&=\Delta\bar{\theta}_{m5}^{67},\qquad b_m=\Delta\bar{\Omega}_{m5}^{67},\qquad 
c_m=\e\Delta\gamma_{m}^{567}.
\end{align}
Exploiting the terms in the square brackets of Eq. \eqref{dI2temp2} gives 
\begin{equation} 
\begin{split}
&\frac{\sin(a+bt)}{b_m^2+c_m^2}\Biggl\{e^{c_m t} [c_m \sin (a_m+b_m t)-b_m \cos (a_m+b_m t)]+b_m \cos (a_m)-c_m \sin(a_m)\Biggl\}+\\
&\frac{\sigma_m\cos(a+bt)}{b_m^2+c_m^2}\Biggl\{e^{c_m t} [c_m \cos(a_m+b_m t)+b_m \sin (a_m+b_m t)]-b_m \sin(a_m)-c_m \cos(a_m)\Biggl\}.\nonumber
\end{split}
\end{equation}
Noticing that $\sigma_m^2=1$, $\cos(\sigma_mx)=\cos(x)$, $ \sin(\sigma_mx)=\sigma_m\sin(x)$ we can collect the first two terms in the two brackets as
\begin{align}
&\gamma_m e^{\gamma_mt}\left[\sin(a+bt)\sin(a_m+b_mt)+\sigma_m\cos(a+bt)\cos(a_m+b_mt)\right]=\nonumber\\
&\gamma_me^{\gamma_mt}\left[\sigma_m^2\sin(a+bt)\sin(a_m+b_mt)+\sigma_m\cos(a+bt)\cos(a_m+b_mt)\right]=\nonumber\\    &\gamma_me^{\gamma_mt}\sigma_m\left[\sin(a+bt)\sin(\sigma_m(a_m+b_mt))+\cos(a+bt)\cos(\sigma_m(a_m+b_mt))\right]=\nonumber\\
&\gamma_me^{\gamma_mt}\sigma_m\cos(a+bt-\sigma_m(a_m+b_mt))\nonumber,
\end{align}
the second two terms in the two brackets as
\begin{align}
&b_me^{\gamma_mt}\left[-\sin(a+bt)\cos(a_m+b_mt)+\sigma_m\cos(a+bt)\sin(a_m+b_mt)\right]=\nonumber\\
&b_me^{\gamma_mt}\left[-\sin(a+bt)\cos(\sigma_m(a_m+b_mt))+\cos(a+bt)\sin(\sigma_m(a_m+b_mt))\right]=\nonumber\\
&-b_me^{\gamma_mt}\left[\sin(a+bt)\cos(\sigma_m(a_m+b_mt))-\cos(a+bt)\sin(\sigma_m(a_m+b_mt))\right]=\nonumber\\
&-b_me^{\gamma_mt}\sin(a+bt-\sigma_m(a_m+b_mt))\nonumber,
\end{align}
the third two terms in the two brackets 
\begin{align}
&b_m\left[\sin(a+bt)\cos(a_m)-\sigma_m\cos(a+bt)\sin(a_m)\right]=\nonumber\\
&b_m\left[\sin(a+bt)\cos(\sigma_ma_m)-\cos(a+bt)\sin(\sigma_ma_m)\right]=\nonumber\\
&b_m\sin(a+bt-\sigma_ma_m),\nonumber
\end{align}
and the last two terms 
\begin{align}
&-\gamma_m\left[\sin(a+bt)\sin(a_m)+\sigma_m\cos(a+bt)\cos(a_m)\right]=\nonumber\\
&-\gamma_m\left[\sigma_m^2\sin(a+bt)\sin(a_m)+\sigma_m\cos(a+bt)\cos(a_m)\right]=\nonumber\\
&-\gamma_m\sigma_m\left[\sin(a+bt)\sin(\sigma_ma_m)+\cos(a+bt)\cos(\sigma_ma_m)\right]=\nonumber\\
&-\gamma_m\sigma_m\cos(a+bt-\sigma_ma_m).\nonumber
\end{align}
We gather all terms and we define $R_{1,m}$ and $R_{2,m}$ as follows
\begin{align}  R_{1,m}&=c_me^{c_mt}\sigma_m\cos(a+bt-\sigma_m(a_m+b_mt))
-c_m\sigma_m\cos(a+bt-\sigma_ma_m)\nonumber\\     &=\mathfrak{Re}\left\{c_me^{c_mt}\sigma_m\exp[i(a+bt-\sigma_m(a_m+b_mt))]-c_m\sigma_m\exp[i(a+bt-\sigma_ma_m)\right\}\nonumber\\
&=\mathfrak{Re}\left\{\sigma_mc_me^{i(a-\sigma_ma_m)}e^{ibt}\left(e^{(c_m-i\sigma_mb_m)t}-1\right)\right\}\nonumber\\
&=\sigma_mc_m\mathfrak{Re}\left\{\left(e^{(c_m-i\sigma_mb_m)t}-1\right)e^{i(a-\sigma_ma_m)}e^{ibt}\right\},\nonumber\\    R_{2,m}&=e^{c_mt}b_m\sin(a+bt-\sigma_m(a_m+b_mt))
-b_m\sin(a+bt-\sigma_ma_m)\nonumber\\
&=\mathfrak{Im}\left\{e^{c_mt}b_m\exp[i(a+bt-\sigma_m(a_m+b_mt))]
-b_m\exp[i(a+bt-\sigma_ma_m)\right\}\nonumber\\
&=\mathfrak{Im}\left\{b_me^{i(a-\sigma_ma_m)}e^{ibt}\left(e^{(c_m-i\sigma_mb_m)t}-1\right)\right\}\nonumber\\
&=b_m\mathfrak{Im}\left\{\left(e^{(c_m-i\sigma_mb_m)t}-1\right)e^{i(a-\sigma_ma_m)}e^{ibt}\right\}.\nonumber
\end{align}
We insert these expressions in Eq. \eqref{dI2temp2} and we return to the variables $\eqref{notation}$ to finally obtain Eq. \eqref{dJ2}.

\section{Appendix C}\label{appD}
The relevant terms to the phase averaging of Eq. \eqref{dJ2} are the exponential terms in $R_{1,m}$ and $R_{2,m}$  involving the angular variables. In particular, if $\Delta\bar{\theta}_{12}^{34}-\sigma_m\Delta\bar{\theta}_{m5}^{67}$ is proportional $\theta$ the integral is zero, and if $\Delta\bar{\theta}_{12}^{34}-\sigma_m\Delta\bar{\theta}_{m5}^{67}=0$, the contribution of the integral is equal to 1. We also note that if $\Delta\bar{\theta}_{12}^{34}-\sigma_m\Delta\bar{\theta}_{m5}^{67}=0$ then $\Delta\bar{\Omega}_{12}^{34}-\sigma_m\Delta\bar{\Omega}_{m5}^{67}=0$ and hence, the phase averaging of Eq. \eqref{dJ2} can be simplified as 
\begin{equation}\label{meanI2}
\frac{\dd \langle J_1^{(2)}\rangle_{\bar{\theta}}}{\dd t}=\sum_{m=1}^4\frac{\mathcal{S}_m}{b_m^2+c_m^2}\left\{\sigma_mc_m\left[e^{c_m t}-\cos(bt)\right]+b_m\sin(bt)\right\},
\end{equation}
where we used again the notation \eqref{notation}. In what follows we will be using the following properties
\begin{equation}
\Delta\bar{\Omega}_{ab}^{cd}=-\Delta\bar{\Omega}_{cd}^{ab},\qquad \Delta\gamma_{b}^{acd}=\Delta\gamma_{a}^{bcd}+2(\gamma_b-\gamma_a).
\end{equation}
We consider the first term $m=1$ and we impose $\Delta\bar{\theta}_{12}^{34}-\sigma_1\Delta\bar{\theta}_{15}^{67}=0 $ which is satisfied when $\mathbf{k}_2=\mathbf{k}_5$, $\mathbf{k}_3=\mathbf{k}_6$, and $\mathbf{k}_4=\mathbf{k}_7$, or when $\mathbf{k}_2=\mathbf{k}_5$, $\mathbf{k}_3=\mathbf{k}_7$, and $\mathbf{k}_4=\mathbf{k}_6$. In both cases the terms on the right hand side of Eq. \eqref{meanI2} are
\begin{align}
b_1&=\Delta\bar{\Omega}_{12}^{34},\qquad 
c_1= \e\Delta\gamma_{1}^{234},\qquad b-\sigma_1b_1=\Delta\bar{\Omega}_{12}^{34}-\Delta\bar{\Omega}_{12}^{34}=0,\nonumber\\
\mathcal{S}_1&=2\sum_{\substack{234\\567}}'T^2_{1234}\frac{\sqrt{\bar{J}_1\bar{J}_2\bar{J}_3\bar{J}_4\bar{J}_2\bar{J}_3\bar{J}_4}}{\sqrt{\bar{J}_1}}e^{\e\Delta\gamma_{1}^{234}t}\delta_{12}^{34}=2\sum_{234}'\bar{J}_2\bar{J}_3\bar{J}_4e^{\e\Delta\gamma_{1}^{234}t}\delta_{12}^{34},\nonumber
\end{align}
which leads to the following term 
\begin{align}
4\sum_{234}'T^2_{1234}\bar{J}_2\bar{J}_3\bar{J}_4\frac{e^{\e\Delta\gamma_{1}^{234}t}\left\{\Delta\bar{\Omega}_{12}^{34}\sin(\Delta\bar{\Omega}_{12}^{34}t)+\e\Delta\gamma_{1}^{234}\left[e^{ \e\Delta\gamma_{1}^{234} t}-\cos(\Delta\bar{\Omega}_{12}^{34}t)\right]\right\}}{(\Delta\bar{\Omega}_{12}^{34})^2+( \e\Delta\gamma_{1}^{234})^2}
\delta_{12}^{34},\nonumber
\end{align}
where the factor 4 is due to the second case $\mathbf{k}_2=\mathbf{k}_5$, $\mathbf{k}_3=\mathbf{k}_7$, and $\mathbf{k}_4=\mathbf{k}_6$. If we neglect the term proportional to $\e$ which leads to an overall $\e^3$ contribution, then the first term of Eq. \eqref{dJ2} is \begin{align}\label{dI2nonconstant}
\frac{\dd \langle J_1^{(2)}\rangle_{\bar{\theta}}}{\dd t}=&4\sum_{234}'T_{1234}^2\bar{J}_2\bar{J}_3\bar{J}_4\frac{ \Delta\bar{\Omega}_{12}^{34}\sin(\Delta\bar{\Omega}_{12}^{34}t)e^{\e\Delta\gamma_{1}^{234}t}}{(\Delta\bar{\Omega}_{12}^{34})^2+(\e p_1)^2}\delta_{12}^{34},
\end{align}
where $p_1$ is given by \eqref{p_m} for $m=1$. Terms for $m=2,3,4$ can be evaluated analogously.

\section{Appendix D}\label{appE}

Equation \eqref{NLSdiss2} can be written as 
\begin{align}\label{db1}
i\frac{\dd b_1}{\dd t}=\epsilon\sum_{234}{}^{'}b_2^*b_3b_4e^{i\Delta\Omega_{12}^{34}t+\epsilon\Delta\gamma_{1}^{234}t}\delta_{12}^{34},
\end{align}
where $b_k=a_k\exp(-i(\Omega_k-i\gamma_k))$. Multiplying the above by $b_1^*$, subtracting its complex conjugate, and taking the mean value leads to
\begin{align}\label{dndt}
\frac{\dd |b_1|^2}{\dd t}=2\epsilon\sum_{234}{}^{'}\mathfrak{Im}\left\{\langle b_1^*b_2^*b_3b_4\rangle e^{i\Delta\Omega_{12}^{34}t+\epsilon\Delta\gamma_{1}^{234}t}\right\}\delta_{12}^{34}.
\end{align}
We consider the time derivative of the four point correlator 
\begin{align}
\frac{\dd}{\dd t}\langle b_1^*b_2^*b_3b_4\rangle=\langle\frac{\dd b_1^*}{\dd t}b_2^*b_3b_4\rangle+\langle\frac{\dd b_2^*}{\dd t}b_1^*b_3b_4\rangle+\langle\frac{\dd b_3}{\dd t}b_1^*b_2^*b_4\rangle+\langle\frac{\dd b_4}{\dd t}b_1^*b_2^*b_3\rangle.
\end{align}
We insert Eq. \eqref{db1} in each term on the right hand side and we integrate both sides between $0$ and $t$ and we obtain
\begin{align}\label{corr4}
\langle b_1^*(t)b_2^*(t)b_3(t)b_4(t)\rangle-\langle \bar{b}_1^*\bar{b}_2^*\bar{b}_3\bar{b}_4\rangle&=i\epsilon\sum_{567}'\langle \bar{b}_5\bar{b}_6^*\bar{b}_7^*\bar{b}_2^*\bar{b}_3\bar{b}_4\rangle\delta_{15}^{67}\int_0^{t}e^{-i\Delta\Omega_{15}^{67}s+\epsilon\Delta\gamma_{1}^{567}s}\dd s\nonumber\\
&+i\epsilon\sum_{567}'\langle \bar{b}_5\bar{b}_6^*\bar{b}_7^*\bar{b}_1^*\bar{b}_3\bar{b}_4\rangle\delta_{25}^{67}\int_0^{t}e^{-i\Delta\Omega_{25}^{67}s+\epsilon\Delta\gamma_{2}^{567}s}\dd s\nonumber\\
&-i\epsilon\sum_{567}'\langle \bar{b}_5^*\bar{b}_6\bar{b}_7\bar{b}_1^*\bar{b}_2^*\bar{b}_4\rangle\delta_{35}^{67}\int_0^{t}e^{i\Delta\Omega_{35}^{67}s+\epsilon\Delta\gamma_{3}^{567}s}\dd s\nonumber\\
&-i\epsilon\sum_{567}'\langle \bar{b}_5^*\bar{b}_6\bar{b}_7\bar{b}_1^*\bar{b}_2^*\bar{b}_3\rangle\delta_{45}^{67}\int_0^{t}e^{i\Delta\Omega_{45}^{67}s+\epsilon\Delta\gamma_{4}^{567}s}\dd s\nonumber\\
&=i\epsilon\sum_{567}'\langle \bar{b}_5\bar{b}_6^*\bar{b}_7^*\bar{b}_2^*\bar{b}_3\bar{b}_4\rangle\delta_{15}^{67}\frac{e^{-i\Delta\Omega_{15}^{67}t+\epsilon\Delta\gamma_{1}^{567}t}-1}{-i\Delta\Omega_{15}^{67}+\epsilon\Delta\gamma_{1}^{567}}\nonumber\\
&+i\epsilon\sum_{567}'\langle \bar{b}_5\bar{b}_6^*\bar{b}_7^*\bar{b}_1^*\bar{b}_3\bar{b}_4\rangle\delta_{25}^{67}\frac{e^{-i\Delta\Omega_{25}^{67}t+\epsilon\Delta\gamma_{2}^{567}t}-1}{-i\Delta\Omega_{25}^{67}+\epsilon\Delta\gamma_{2}^{567}}\nonumber\\
&-i\epsilon\sum_{567}'\langle \bar{b}_5^*\bar{b}_6\bar{b}_7\bar{b}_1^*\bar{b}_2^*\bar{b}_4\rangle\delta_{35}^{67}\frac{e^{i\Delta\Omega_{35}^{67}t+\epsilon\Delta\gamma_{3}^{567}t}-1}{i\Delta\Omega_{35}^{67}+\epsilon\Delta\gamma_{3}^{567}}\nonumber\\
&-i\epsilon\sum_{567}'\langle \bar{b}_5^*\bar{b}_6\bar{b}_7\bar{b}_1^*\bar{b}_2^*\bar{b}_3\rangle\delta_{45}^{67}\frac{e^{i\Delta\Omega_{45}^{67}t+\epsilon\Delta\gamma_{4}^{567}t}-1}{i\Delta\Omega_{45}^{67}+\epsilon\Delta\gamma_{4}^{567}},
\end{align}
where we assumed that the six points correlator does not evolve in time and therefore all terms in the six points correlators are equal to their value at initial time $\bar{b}_{\mathbf{k}}\equiv b_{\mathbf{k}}(0)$. This corresponds to the condition of closure
\begin{align*}
    \frac{\dd \langle b_j^*b_6^*b_7^*b_3b_4b_5\rangle}{\dd t}=\mathcal{O}(\epsilon),\quad j=1,2\qquad \frac{\dd \langle b_1^*b_2^*b_5^*b_ib_6b_7\rangle}{\dd t}=\mathcal{O}(\epsilon),\quad i=3,4. 
\end{align*}
We now insert Eq. \eqref{corr4} into Eq. \eqref{dndt} and we obtain
\begin{align}\label{all6}
\frac{\dd |b_1|^2}{\dd t}=2\epsilon^2\sum_{\substack{234\\567}}'\mathfrak{Im}&\left\{\langle \bar{b}_1^*\bar{b}_2^*\bar{b}_3\bar{b}_4\rangle e^{i\Delta\Omega_{12}^{34}t+\epsilon\Delta\gamma_{1}^{234}t}\right.\nonumber\\
&+i\langle \bar{b}_5\bar{b}_6^*\bar{b}_7^*\bar{b}_2^*\bar{b}_3\bar{b}_4\rangle\delta_{15}^{67}\frac{e^{-i\Delta\Omega_{15}^{67}t+\epsilon\Delta\gamma_{1}^{567}t}-1}{-i\Delta\Omega_{15}^{67}+\epsilon\Delta\gamma_{1}^{567}}e^{i\Delta\Omega_{12}^{34}t+\epsilon\Delta\gamma_{1}^{234}t}\nonumber\\
&+i\langle \bar{b}_5\bar{b}_6^*\bar{b}_7^*\bar{b}_1^*\bar{b}_3\bar{b}_4\rangle\delta_{25}^{67}\frac{e^{-i\Delta\Omega_{25}^{67}t+\epsilon\Delta\gamma_{2}^{567}t}-1}{-i\Delta\Omega_{25}^{67}+\epsilon\Delta\gamma_{2}^{567}}e^{i\Delta\Omega_{12}^{34}t+\epsilon\Delta\gamma_{1}^{234}t}\nonumber\\
&-i\langle \bar{b}_5^*\bar{b}_6\bar{b}_7\bar{b}_1^*\bar{b}_2^*\bar{b}_4\rangle\delta_{35}^{67}\frac{e^{i\Delta\Omega_{35}^{67}t+\epsilon\Delta\gamma_{3}^{567}t}-1}{i\Delta\Omega_{35}^{67}+\epsilon\Delta\gamma_{3}^{567}}e^{i\Delta\Omega_{12}^{34}t+\epsilon\Delta\gamma_{1}^{234}t}\nonumber\\
&\left.-i\langle \bar{b}_5^*\bar{b}_6\bar{b}_7\bar{b}_1^*\bar{b}_2^*\bar{b}_3\rangle\delta_{45}^{67}\frac{e^{i\Delta\Omega_{45}^{67}t+\epsilon\Delta\gamma_{4}^{567}t}-1}{i\Delta\Omega_{45}^{67}+\epsilon\Delta\gamma_{4}^{567}} e^{i\Delta\Omega_{12}^{34}t+\epsilon\Delta\gamma_{1}^{234}t}\right\}\delta_{12}^{34}.
\end{align}
The first term in the curly brackets can be evaluated by using Wick's decomposition
\begin{align*}
\langle b_1^*b_2^*b_3b_4\rangle=|b_3|^2|b_4|^2(\delta_3^1\delta_4^2+\delta_3^2\delta_4^1),
\end{align*}
so that 
\begin{align*}
\langle \bar{b}_1^*\bar{b}_2^*\bar{b}_3\bar{b}_4\rangle=|\bar{b}_3|^2|\bar{b}_4|^2(\delta_3^1\delta_4^2+\delta_3^2\delta_4^1)e^{i\Delta\Omega_{12}^{34}t+\epsilon\Delta\gamma_{1}^{234}t}=|\bar{b}_1|^2|\bar{b}_2|^2 e^{-2\epsilon \gamma_2 t},
\end{align*}
which we note is a pure real quantity and hence, it will not give contribution to Eq. \eqref{dndt}. The other terms in the curly brackets can be evaluated again by using Wick's decomposition: for instance, for the second term in Eq. \eqref{all6} we can use
\begin{align*}
    \langle b_3b_4b_5b_2^*b_6^*b_7^*\rangle=|b_2|^2|b_6|^2|b_7|^2\left(\delta_{3}^{2}\delta_{4}^{6}\delta_{5}^{7}+\delta_{3}^{6}\delta_{4}^{7}\delta_{5}^{2}+\delta_{3}^{7}\delta_{4}^{2}\delta_{5}^{6}+\delta_{3}^{2}\delta_{4}^{7}\delta_{5}^{6}+\delta_{3}^{7}\delta_{4}^{6}\delta_{5}^{2}+\delta_{3}^{6}\delta_{4}^{2}\delta_{5}^{7}\right),
\end{align*}
and the non trivial terms are those proportional to $\delta_{3}^{6}\delta_{4}^{7}\delta_{5}^{2}$ and  $\delta_{3}^{7}\delta_{4}^{6}\delta_{5}^{2}$. Evaluating both terms leads to
\begin{align}
2i|\bar{b}_2|^2|\bar{b}_3|^2|\bar{b}_4|^2\frac{e^{-i\Delta\Omega_{12}^{34}t+\epsilon\Delta\gamma_{1}^{234}t}-1}{-i\Delta\Omega_{12}^{34}+\epsilon\Delta\gamma_{1}^{234}} e^{i\Delta\Omega_{12}^{34}t+\epsilon\Delta\gamma_{1}^{234}t}\delta_{12}^{34}.
\end{align}
Rationalizing the denominator and removing the term proportional to $\epsilon$, which gives an overall $\epsilon^3$ contribution, leads to
\begin{align}
-2|\bar{b}_2|^2|\bar{b}_3|^2|\bar{b}_4|^2\frac{\Delta\Omega_{12}^{34}\left(e^{-i\Delta\Omega_{12}^{34}t+\epsilon\Delta\gamma_{1}^{234}t}-1\right)}{\left(\Delta\Omega_{12}^{34}\right)^2+\left(\epsilon\Delta\gamma_{1}^{234}\right)^2} e^{i\Delta\Omega_{12}^{34}t+\epsilon\Delta\gamma_{1}^{234}t}\delta_{12}^{34}.
\end{align}
Carrying out the multiplication gives
\begin{align*}
-2|\bar{b}_2|^2|\bar{b}_3|^2|\bar{b}_4|^2\frac{\Delta\Omega_{12}^{34}\left(e^{2\epsilon\Delta\gamma_{1}^{234}t}-e^{\epsilon\Delta\gamma_1^{234}t}\cos\left(\Delta\Omega_{12}^{34}t\right)-i\sin\left(\Delta\Omega_{12}^{34}t\right)e^{\epsilon\Delta\gamma_{1}^{234}t}\right)}{\left(\Delta\Omega_{12}^{34}\right)^2+\left(\epsilon\Delta\gamma_{1}^{234}\right)^2}\delta_{12}^{34}.
\end{align*}
According to Eq. \eqref{dndt} we only need to consider the imaginary part so we have 
\begin{align*}
2|\bar{b}_2|^2|\bar{b}_3|^2|\bar{b}_4|^2\frac{\Delta\Omega_{12}^{34}\sin\left(\Delta\Omega_{12}^{34}t\right)e^{\epsilon\Delta\gamma_{1}^{234}t}}{\left(\Delta\Omega_{12}^{34}\right)^2+\left(\epsilon\Delta\gamma_{1}^{234}\right)^2}\delta_{12}^{34}.
\end{align*}
The calculation for the other terms can be done in analogous way and, after inserting all terms into Eq. \eqref{all6}, we recover Eq. \eqref{dI2nonconstantapp} for $m=1$.

\bibliography{biblio}%

\end{document}